\documentclass[fleqn]{2017SCGE}
\usepackage{bm,color}

\def\aap{A\&A}
\def\aa{Astron. Astrophys.}

\def\apj{Astrophys. J.}
\def\apjl{Astrophysics J. Lett.}
\def\apjsup{Astrophys. J. Suppl. Ser.}

\def\pasj{Publ. Astron. Soc. Jap.}

\def\sp{Solar Physics}

\begin{document}
\ensubject{subject}

\ArticleType{Article}
\ReceiveDate{}
\AcceptDate{}

\title{Diagnostic  of Spectral Lines in Magnetized Solar Atmosphere: 
Formation  of the H$\beta$  Line in Sunspots}{Formation of H$\beta$ Line in Solar Magnetic Atmospheres}
\author[]{Hongqi Zhang}{{hzhang@bao.ac.cn}}

\AuthorMark{Hongqi Zhang}

\address[]{Key Laboratory of Solar Activity, National Astronomical Observatories of the Chinese Academy of Sciences, 100101, Beijing China}

\abstract{ Formation of the H$\beta$ $\lambda$4861.34 \AA{} line is an important topic related to the diagnosis of the basic configuration of magnetic fields in the solar and stellar chromospheres.  Specifically, broadening of the H$\beta$ $\lambda$4861.34 \AA{} line occurs due to the magnetic and micro-electric fields in the solar atmosphere.  The formation of H$\beta$ in the  model umbral atmosphere is presented based on the assumption of non-local thermodynamic equilibrium. It is found that the model umbral chromosphere is transparent to the Stokes parameters of the H$\beta$ line, which implies that the observed signals of magnetic fields at sunspot umbrae via the H$\beta$ line originate from the deep solar atmosphere, where $\lg \tau_c\approx-1$ (about 300 km in the photospheric layer for our calculations). This is in contrast to the observed Stokes signals from non-sunspot areas, which are thought to primarily form in the solar chromosphere.       }


\maketitle
\begin{multicols}{2}  

\section{Introduction}

 With the discovery of the Zeeman effect by P. Zeeman (1897)\cite{Zeeman1897}, scientists can therefore extract information on magnetic fields in remote objects through spectro-polarimetry. Hale (1908)\cite{Hale1908} was the first to make use of this principle in astrophysics, which led him to find pronounced Zeeman splitting in sunspots. Quantitative measurements of solar magnetic fields were made possible. Since the 1980s, a modern solar vector magnetograph (Solar Magnetic Field Telescope) was developed in China (Ai and Hu, 1986)\cite{Ai86}. It was the inauguration of the systematic observations and the corresponding theoretical research of solar magnetic fields in China. 

Chromosphere is a very important layer in the diagnostic of solar and stellar activities\cite{Zirin88}. The H$\beta$ $\lambda$4861.34 \AA{} line has been used at Huairou Solar Observing Station (HSOS) of National Astronomical Observatories of the Chinese Academy of Sciences for the measurements of the solar chromospheric magnetic fields (Zhang \& Ai, 1987)\cite{ZhangA87}. Similar magnetograms  were obtained at other observatories, such as Crimea and Kitt Peak  \cite{SeBu8,Tsap71,Giovanelli80}. 
These observations have proved useful to enhance our understanding of the different properties of solar active phenomena caused by magnetic fields \cite{Georg19}.

Diagnostic on the formation layers of spectral lines in solar or stellar atmospheres is a basic topic, especially for the measurements of the magnetic fields with Stokes parameters of spectral lines. 
In the following, we present the analysis on the formation of the H$\beta$ $\lambda$4861.34 \AA{} line in the magnetic fields of solar atmosphere, especially in the sunspot umbral atmosphere.

\section{ Solar Model Atmospheres }

The H$\beta$ line forms in a relatively wide solar  atmosphere. This is related to the basic states on the statistical distribution of energy level populations of hydrogen atoms in the solar atmosphere.
To study  the formation of the H$\beta$ line in the magnetized solar atmosphere, 
the non-local thermodynamic equilibrium (NLTE) population departure coefficients $b_i$ are defined as:
\begin{equation}
\label{eq:souf1}
b_l=N_l/N^{LTE}_l, ~~~ b_u=N_u/N^{LTE}_u,
\end{equation}
where $N_{(l,u)}$ is the actual population and $N^{LTE}_{(l,u)}$ the Saha-Boltzmann values for the lower ($l$) and upper ($u$) levels.  The treatment of NLTE is suitable  for analyzing the formation of spectral lines in the upper solar atmosphere due to the weakening contribution of particle collisions,  while the approximation of local thermodynamic equilibrium (LTE) is normally acceptable in the lower solar atmosphere due to the high density  of the particle populations, where the departure coefficients are $b_i\approx1$ in eq. (\ref{eq:souf1}). 

\begin{figure}[H] 
\vspace{0.2cm}
{\includegraphics[width=75mm]{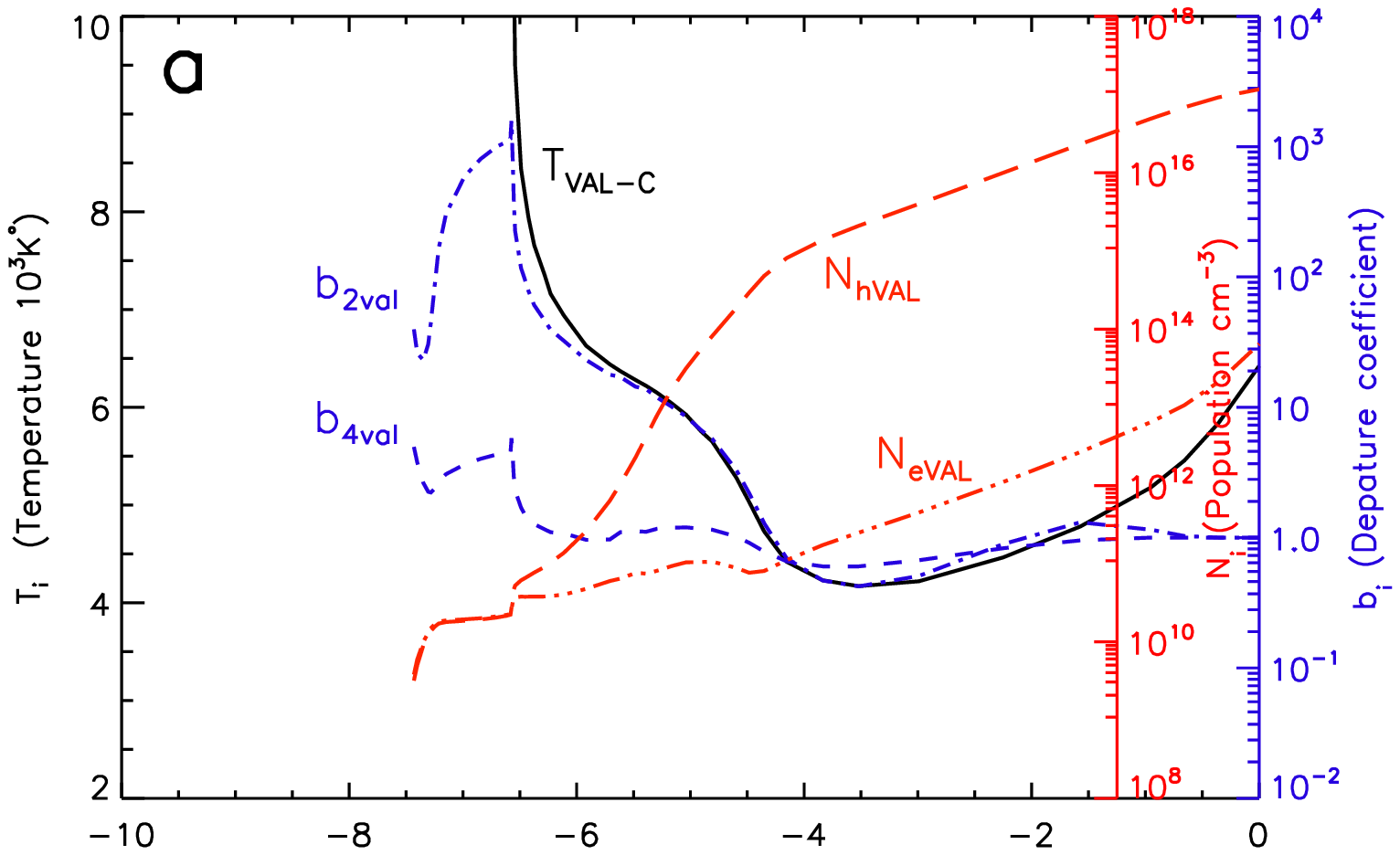}\vspace{-0.001cm}
 \includegraphics[width=75mm]{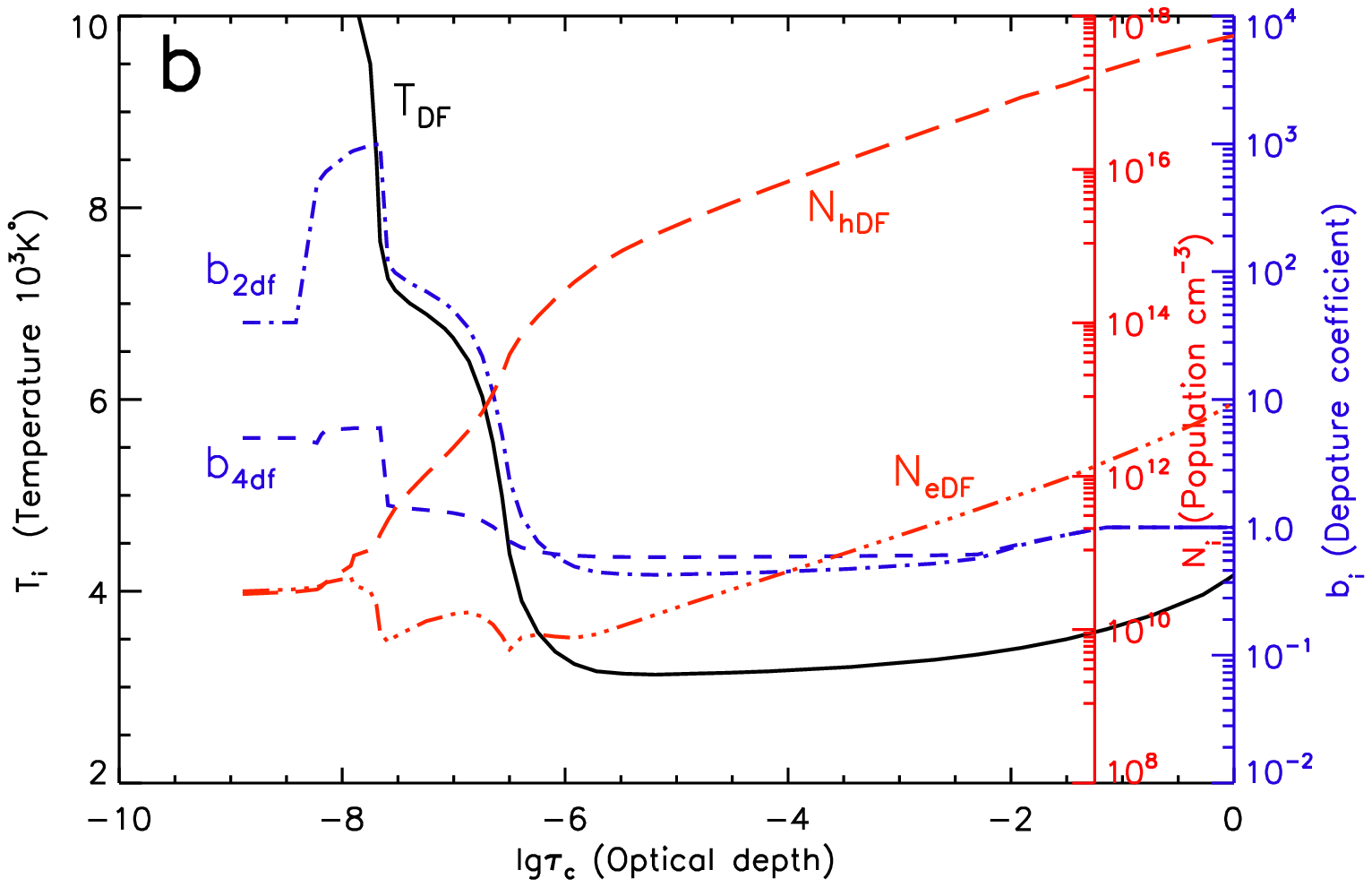}}
\vspace{-0.2cm}
\caption{Distribution of  ${T}$,  ${N_h}$,  ${N_e}$, and the departure coefficients (${b_2}$ and ${b_4}$) of the hydrogen levels. The continuum optical depth is displayed in 
$\lg \tau_c$ scale in the atmospheric model of the quiet Sun (a) by Vernazza, Avrett \& Loeser (VAL, 1981)\cite{Vernazza81} and the sunspot umbra model (b) by Ding \& Fang (DF, 1991)\cite{Ding91}. 
}
\label{fig:departureb}
\end{figure}

Figure \ref{fig:departureb}a shows the atmospheric model of the quiet Sun with the distributions of the departure coefficients $b_2$ and $b_4$  of the hydrogen atomic levels $n\!=\!2$ and $\!n=\!4$  at different continuum optical depths (Vernazza, Avrett \& Loeser, 1981\cite{Vernazza81}).  It is found that the  departure coefficients $b_i$ in the lower solar atmosphere are of the order unity, and they increase drastically 
as hydrogen density ${N_h}$ and electron density ${N_e}$ decrease, and   
as the temperature increases at low optical depths. The departure coefficients tend to decrease at very thin depths.

A sunspot umbral model with the distributions of temperature ${T}$,  hydrogen density ${N_h}$ and electron density ${N_e}$ in Figure \ref{fig:departureb}b  was provided by Ding \& Fang (1991)\cite{Ding91}.

Assuming that the same thermodynamic mechanism operates both in the quiet Sun and sunspot umbrae models, the  departure coefficients $b_i$ of the hydroden atomic levels in the  latter atmosphere model can be phenomenologically provided by using the tendencies of the departure coefficients $b_i$ found for the quiet Sun as a proxy. 
Figure \ref{fig:departureb}b shows the population departure coefficient $b_i\approx1$ in the deep umbral model atmosphere, which increases with height  in the interval $\lg \tau=[ -6, -7.5]$ and  decreases gradually with height  in the range $\lg \tau=[ -7.5, -9]$ .

In thermodynamic equilibrium, atoms are distributed over their bound levels according to the Boltzmann excitation equation \cite{Mihalas78}. The continuum normally forms in the lower solar atmosphere, thus
the Planck function $B(T,\nu)$ can be used as its source function   
\begin{equation}
\label{eq:souf2}
B^C(T,\nu)=\frac{2h\nu^3}{c^2}\frac{1}{\displaystyle\exp(h\nu/kT)-1},
\end{equation} 
and the source function of  H$\beta$ line is
\begin{equation}
\label{eq:souf3}
S^L(H_\beta)=\frac{2h\nu^3}{c^2}\frac{1}{\displaystyle\frac{b_2}{b_4}\exp(h\nu/kT)-1},
\end{equation} 
where the values of $b_2$ and $b_4$ are shown  in Figure \ref{fig:departureb}.  
The self-consistency of these departure coefficients of the H$\beta$ line with the observations  can be found in the following sections. 

\section{Hydrogen Lines, Magnetic Fields and Ionized Solar Atmosphere}

To analyze the contribution of the magnetic and micro-electric fields to the broadening of hydrogen lines in the solar atmosphere, the  Hamiltonian can be written in the following condensed form
\begin{equation}
\label{eq:hami1}
 \mathcal{H}   =\mathcal{H}_0+\mathcal{H}_Z+\mathcal{H}_S,  
\end{equation}
where $\mathcal{H}_0$ includes the Hamiltonian $\mathcal{H}_{NR}$, $\mathcal{H}_{fine}$ and $\mathcal{H}_{hyper} $, and $ \mathcal{H}_Z=\mu_0\textbf{ B}_e \cdot (\textbf{L}+2\textbf{S})$ and $ \mathcal{H}_S=e_0\textbf{ E}_e \cdot \textbf{r}$. We have purposely chosen the Hamiltonian $ \mathcal{H}_Z$ and $ \mathcal{H}_S$ of the Zeeman and Stark effects to analyze the broadening of hydrogen lines (such as, the H$\beta$ line) in the solar atmosphere, as magnetic and micro-electric fields commonly contribute to these lines.   The general results for the fine structure of hydrogen lines can be found in the monograph of Bethe \& Salpeter (1957)\cite{Bethe57}. 

Next, we will consider  the hydrogen atoms and  perturbing, charged particles in external magnetic and electric fields (eq. \ref{eq:hami1}).   The  perturbation equation is
\begin{equation}
 \label{eq:HB12}
\mathcal{H}_0+\mathcal{H}_Z+\mathcal{H}_S\mid\psi \rangle=(E_0+E' )\mid\psi \rangle,
\end{equation}
where $E_0$ and $E'$ are the energy's eigenvalue and perturbation, respectively. 
One can calculate the shifts $\triangle\lambda_j$ of the sub-components of the spectral line by means of eq. (\ref{eq:HB12}) when the magnetic and electric fields are considered.   

Left-multiply eq. (\ref{eq:HB12}) by $\langle\psi|$, and use the identity for the perturbing Hamiltonian 
\begin{equation}
\label{eq:HB17}
\langle n,l,j,m_j\mid \mathcal{H}_Z+\mathcal{H}_S\mid n,l',j',m_j'\rangle \equiv K_{q,q'},
\end{equation}
the suffixes $q$ and $q'$ in $K$ refer to various possible wave function states. 
Assume the perturbed wave function to be a linear combination of unperturbed wave functions:
\begin{equation}
\label{eq:HB18}
\mid\psi\rangle=\sum_{q'=1}^{2n^2}c_{q'}\mid\psi_{q'} \rangle,
\end{equation}
the eq. (\ref{eq:HB17}) then becomes
\begin{equation}
\sum_{q'=1}^{2n^2}(K_{q,q'}-E'\delta_{qq'})c_{q'} =0,
\label{eq:HB19}
\end{equation}
where $q=1,2,....,2n^2$. Normally, the shifts of the energy levels depend  on the distribution of the magnetic and electric field nearby the hydrogen atom. 

Normally, the general solution of the eigenvalue problem for hydrogen lines relates to the direction of the magnetic field and microscopic electric field. By means of eq. (\ref{eq:HB19}), we can find the $2n^2$ possible splits of the upper and lower energy levels and the corresponding $2n^2$ probabilities of the wave functions in various directions and intensities of the electric field for each sub-level of the hydrogen line. Then, we can calculate the shifts $\triangle\lambda_j$ of the sub-components $j$ of the spectral line (Zhang, 1987)\cite{ZhangA87}. 
The calculation of the Hermitian matrix of the perturbing Hamiltonian was presented also by Casini and Landi Degl'Innocenti (1993)\cite{Casini93}. 
 
As  the microscopic electric field in the solar atmosphere is statistically isotropic, we assumed here that the polarized states of the sub-components of the H$\beta$ line still rely on the magnetic field. Because the perturbed wave function is a linear combination of unperturbed wave functions, the polarized states of the sub-components of the line are also related to the magnetic quantum number $m$, e.g., $\triangle m=0$ for the $\pi$ component and $\triangle m=\pm 1$ for both $\sigma$ components.

\begin{figure}[H] 
\hspace{-0.2cm}
\centerline{\includegraphics[width=90mm]{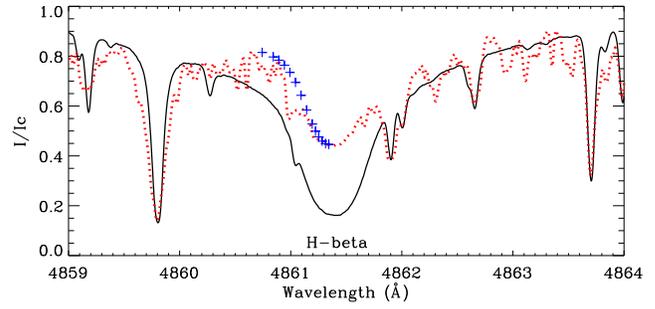}}
\vspace{-0.5cm}
\caption{ H$\beta$ line profiles of the quiet Sun (solid line) and of the  sunspot umbra  (red dotted line)  observed by the National Solar Observatory of USA (L. Wallace, K. Hinkle, and W. C. Livingston, http://diglib.nso.edu/ftp.html).  The blue plus symbols mark the results of the numerical calculation.}
\label{fig:StokesHV2}
\end{figure}
  
The observed  profiles of the H$\beta$ $\lambda$4861.34 \AA{} line in the quiet Sun and sunspot umbra with the overlap of some photospheric lines in the wings are shown in Figure \ref{fig:StokesHV2}.
In the following, we also neglect the asymmetry of the spectral line, which is normally believed to be caused by the gradient of the line of sight velocity field in the solar atmosphere\cite{Almeida92}.

\section{Formation of the H$\beta$ Line in Solar Magnetic Atmosphere} 

The numerical Stokes profiles ($I$, $Q$, $U$, $V$) of the H$\beta$ line can be calculated by means of the Unno-Rachkovsky equations of polarized radiative transfer of spectral lines 
(Unno (1956) \cite{Unno56} and Rachkovsky (1962a, b) \cite{Rachkovsky62a,Rachkovsky62b})
\begin{equation}
\label{radtrf}
\mu\frac{d}{d\tau_c}\left(\begin{array}{c}
I\\
Q\\
U\\
V
\end{array} \right)=
\left(\begin{array}{cccc}
\eta_{0}+\eta_{I} & \eta_{Q} & \eta_{U} & \eta_{V}\\
\eta_{Q} & \eta_{0}+\eta_{I} &  \rho_{V} & -\rho_{U}\\
\eta_{U} & -\rho_{V} & \eta_{0}+\eta_{I} & \rho_{Q}\\
\eta_{V} & \rho_{U} & -\rho_{Q} & \eta_{0}+\eta_{I}
\end{array} \right)
\left( \begin{array}{c}
I-S\\
Q\\
U\\
V
\end{array} \right)
\end{equation}
where the symbols have their usual meanings as given by Landi Degl’Innocenti (1976) \cite{Land76}, $d\tau_c\!=\!-\kappa_cds$,  $\eta_I$, $\eta_Q$, $\eta_U$, $\eta_V$ are the Stokes absorption coefficient parameters and $\rho_Q$, $\rho_U$ and $\rho_V$ are related to magneto-optical effects,  { $S$ represents the source functions related to eq. (\ref{eq:souf2}) for the continuum and eq. (\ref{eq:souf3}) for the H$\beta$ line.  }     

By comparing with the formulae of the non-polarized hydrogen line proposed by Zelenka (1975)\cite{Zelenka75}, the contributions of the statistically distributed microscopic electric field with the Zeeman effects in eqs. (\ref{eq:HB17}-\ref{eq:HB19}) rising from the magnetic field in the solar atmosphere are considered in the  absorption matrix of eq. (\ref{radtrf}) for the H$\beta$ line (see, \cite{ZhangA87,Zhang19}).  

It is  assumed here that, when the degeneracy of an energy level disappears under the actions of magnetic  and microscopic electric fields,  each magnetic energy level keeps its original departure coefficient.

\begin{figure}[H]  
\begin{center}
\includegraphics[width=70mm]{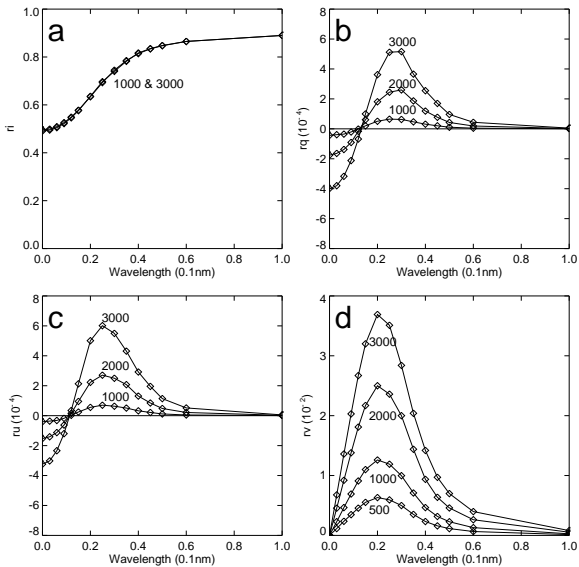}  
\end{center}
\caption{Profiles of Stokes parameters $r_i,r_q,r_u$, and $r_v$ (i.e. $I/I_c$, $Q/I_c$, $U/I_c$  and $V/I_c$) of the H$\beta$ line calculated for  the umbral  atmospheric model  by Ding \& Fang (DF, 1991)\cite{Ding91}. The homogeneous magnetic field intensity of 500, 1000, 2000, 3000G for an inclination of $\psi$=30$^\circ$, an azimuth of $\varphi$=22.5$^\circ$ and $\mu =1$. $I_c$ is the continuum.
\label{fig:HbumbStok}}
\end{figure}

In Figure \ref{fig:HbumbStok}, it is noticed that the numerically calculated residual intensity  at the core of the Stokes $I$ profile is of the order 0.5  in the umbral model atmosphere, and the profile of Stokes $I$ is roughly consistent with the observed umbral one in Figure \ref{fig:StokesHV2}, if the blended lines are ignored. (The calculated result is also marked by plus symbols in the blue wing of the line in Figure \ref{fig:StokesHV2}.) Note that it is weaker than that of the quiet Sun. There are no obvious differences in the Stokes $I$ values between the umbral magnetic field strengths of 1000 and 3000 G. The amplitude of the Stokes $V$ is of the order 10$^{-2}$, and those of Stokes $Q$ and $U$ are $10^{-4}$. 
These values are similar to those of the quiet Sun (see Figure 11 of Zhang, 2019 \cite{Zhang19}). This means measuring the transverse components of the magnetic fields of sunspots is still a technological challenge due to their very weak signals. 
 
Moreover, the peak values of Stokes $V$ are located at 0.2 \AA{} from the line center as calculated using the umbral model atmosphere in Figure \ref{fig:HbumbStok}, and at 0.3 \AA{} for the quiet Sun (see Figure 11 of Zhang, 2019 \cite{Zhang19}). Similar tendencies for Stokes $Q$ and $U$ can be found as well. This reflects that the contribution of the Holtsmark broadening in the umbral atmosphere, which depends on the density of the charged particles,  is relative weaker than that in the quiet Sun. This result is consistent with the abundance of the electrons $N_e$ in both models.  
 
The key is to know where in the solar atmosphere a given spectral line is formed, which can be used to estimate the possible formation height of the observed magnetogram. This information is provided by the contribution function $C_\textbf{  I}$, such as defined by Jin (1981) \cite{Jin81} and Stenflo (1994) \cite{Stenflo94}, for the Stokes vector $\textbf{  I}=(I,Q,U,V)$\cite{Landi04}  
\begin{equation}
\label{eq:contri}
\textbf{  I}(0)=\int _0 ^{\infty}C_\textbf{  I}(\tau_c)d\tau_c=\int _{-\infty}^{\infty}C_\textbf{  I}(x)dx,
\end{equation}
where $C_\textbf{ I}(x)=(\textrm{ln}10)\tau_c C_\textbf{  I}(\tau_c)$ and $x=\textrm{lg}\tau_c$,  $\tau_c$ is the continuum optical depth at 5000 \AA. Equation (\ref{eq:contri}) provides the contribution to the emergent Stokes parameters from the different layers of the solar atmosphere with the equivalent source functions $S_\textbf{  I}^\star$ (Zhang, 1986)\cite{Zhang86}. 
 
 The contribution functions of the H$\beta$ line in the magnetic atmospheric model of the quiet Sun were presented by Zhang and Zhang (2000)\cite{{Zhang00}} and Zhang (2019)\cite{Zhang19}. 
The emergent Stokes parameters at the H$\beta$ line center in the quiet Sun almost form in the middle chromosphere (1500 --1600 km), and those in the blue wing, -0.45\AA{} from the H$\beta$ line center, form in the photosphere  (300 km)\cite{ZhangA87}.   

\begin{figure}[H]  
\begin{center}
\includegraphics[width=70mm]{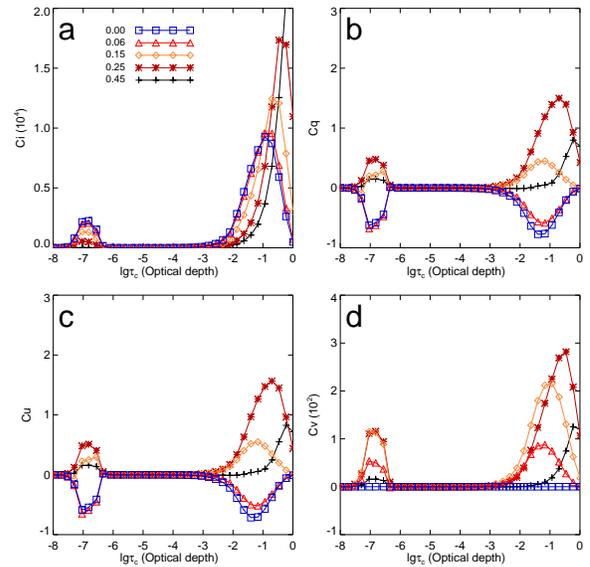}  
\end{center}
\caption{Contribution functions $C_i,C_q,C_u$, and $C_v$ of Stokes parameters $I$, $Q$, $U$, and $V$ of   the numerically calculated H$\beta \lambda$4861.34 \AA{} line for the umbral  atmospheric model by Ding \& Fang (DF, 1991)\cite{Ding91} at  wavelengths of $\triangle\lambda$=0.45 (pluses), 0.25 (asterisks),  0.15 (diamonds), 0.06 (triangles) and 0.0  (squares) \AA{} from the H$\beta$ line center. B=1000 G, $\psi$=30$^\circ$,  $\varphi$=22.5$^\circ$ and $\mu =1$. $\tau_c$ is continuum optical depth at 5000 \AA{}. The horizontal coordinates are in logarithmic scale.
\label{fig:umbralcontrib}}
\end{figure}
   
Figure \ref{fig:umbralcontrib} shows the contributions of the Stokes parameter $I$, $Q$, $U$, and $V$ in the umbra model at 0.0, 0.06, 0.15, 0.25, and 0.45 \AA{} from the center of the H$\beta$ line.  It is noticed that  the contributions of the emergent Stokes parameters in the higher solar atmosphere (the continuum optical depth $\lg \tau_c\!\approx\!-7$) are much weaker than that in the lower atmosphere ($\lg \tau_c\!\approx\!-1$), even though there are some differences in the Stokes parameters at different wavelengths from the center of the H$\beta$ line. This result implies that the upper umbral atmosphere is  transparent to the H$\beta$ line. The  signals of the emergent Stokes parameters of the H$\beta$ line in the umbrae  formed mainly in the lower solar atmosphere, while those near the line center in the quiet Sun form in the chromosphere (see Figure 12 of Zhang, 2019 \cite{Zhang19}).  This result can also be found by comparing the residual intensities at the center of the H$\beta$ line for the quiet Sun and the umbra in Figure \ref{fig:StokesHV2} due to different absorptions of the line.

A similar observational result of the sunspots in the white light has been named as the Wilson effect. Bray and Loughhead (1965)\cite{Bray65} contended that the true explanation of the Wilson effect lies in the higher transparency of the spot material compared to the photosphere. 
The morphological evidence can be found in Figure 13 of Zhang (2019)\cite{Zhang19}. Similar patterns of a sunspot umbra occur both in the photospheric and the H$\beta$ filtergrams, even when the bright flare ribbons in H$\beta$ are attached to the umbrae in the H$\beta$ filtergram.
The  blended lines from the deep layer cause the reversed signals in the umbral areas of the H$\beta$ magnetograms as indicated by Zhang (2019)\cite{Zhang19,Zhang93}.  
 
\begin{figure}[H]  
\begin{center}
\includegraphics[width=85mm]{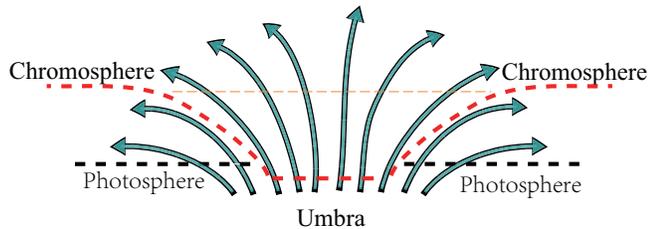}  
\end{center}
\vspace{-0.5cm}
\caption{A schematic sketch  of the estimated formation height of the measured Stokes parameters of the  H$\beta$ line for sunspot regions. The red thick dashed line marks the major contribution heights of the Stokes parameters of the H$\beta$ line, while the orange thin dashed line shows the weak contribution of the line in the chromosphere. 
\label{fig:armodel}}
\end{figure}
  
A simple schematic sketch of the formation heights of the Stokes parameters observed by the H$\beta$ line is proposed in Figure \ref{fig:armodel}.  The red thick dashed line shows the rough detectable heights in the H$\beta$ magnetograms. In the umbrae, the observed Stokes parameters mainly form in deep layers in Figure \ref{fig:umbralcontrib}, although weak contributions also arise from the chromosphere in Figure \ref{fig:armodel}.
This means that the H$\beta$ line in the sunspot umbrae forms at the continuum optical depth $\lg \tau_c\!\approx\!-1$ for our calculations, i.e., about 300 km in the photospheric layer.              

\section{Discussions and Summary}

In this paper,  the 
departure coefficients of the hydrogen atom in the model umbral atmosphere under the NLTE hypothesis were presented and compared with their distributions in the quiet Sun.  
The locations of the peak values of the Stokes $V$ of the H$\beta$ line are close to the line center in the model umbral atmosphere as compared to those of the quiet Sun. This result is consistent with the weaker contribution of the micro-electric field to broadening the H$\beta$ line  in the solar umbral atmosphere than in the quiet Sun. 

It is worth noting that, because the sunspot umbra exhibits lower temperature characteristics compared to the quiet Sun, the umbral chromosphere is generally transparent to the spectral lines that require higher temperature excitation, such as the H$\beta$ line. 
Moreover, the  ratio of the emergent Stokes parameters of spectral lines from different depths of the solar atmosphere depends on the the parameters of the magnetic and atmospheric models used in the radiative transfer equations. In order to verify it,  accurate observations are still necessary and important.  
  
\section{Acknowledgements}

The author would like to thank the  referees for their comments and suggestions, which improved the paper. 
This study is supported by grants from the National Natural Science Foundation of China (NSFC 11673033, 11427803, 11427901) and by Huairou Solar Observing Station, Chinese Academy of Sciences.

\end{multicols}
\end{document}